\documentclass[pra,aps,nopacs,superscriptaddress,nofootinbib]{revtex4}
\usepackage{graphicx}
\usepackage{dcolumn}
\usepackage{bm}
\usepackage{amsmath}
\usepackage{epsfig}

\def\rb{{\bf r}}  \def\jb{{\bf j}}
\def\Te{{T_e}}  \def\Tl{{T_l}}
\def\Ce{{c_e}}  \def\Cl{{c_l}}
\def\ke{{\kappa_e}}  \def\kl{{\kappa_l}}

\begin{document}

\title{Optical generation of intense ultrashort magnetic pulses at the nanoscale}
\author{Anagnostis~Tsiatmas}
\affiliation{Optoelectronics Research Centre and Centre for Photonic Metamaterials, University of Southampton, Southampton SO17 1BJ, United Kingdom}
\author{Evangelos~Atmatzakis}
\affiliation{Optoelectronics Research Centre and Centre for Photonic Metamaterials, University of Southampton, Southampton SO17 1BJ, United Kingdom}
\author{Nikitas~Papasimakis}
\affiliation{Optoelectronics Research Centre and Centre for Photonic Metamaterials, University of Southampton, Southampton SO17 1BJ, United Kingdom}
\author{Vassili~Fedotov}
\affiliation{Optoelectronics Research Centre and Centre for Photonic Metamaterials, University of Southampton, Southampton SO17 1BJ, United Kingdom}
\author{Boris Luk'yanchuk}
\affiliation{Data Storage Institute, Agency for Science, Technology and Research, Singapore 117608}
\author{Nikolay~I.~Zheludev}
\affiliation{Optoelectronics Research Centre and Centre for Photonic Metamaterials, University of Southampton, Southampton SO17 1BJ, United Kingdom}
\affiliation{Centre for Disruptive Photonic Technologies, Nanyang Technological University, Singapore 637371}
\author{F.~Javier~Garc\'{\i}a~de~Abajo}
\email{J.G.deAbajo@nanophotonics.es}
\affiliation{Optoelectronics Research Centre and Centre for Photonic Metamaterials, University of Southampton, Southampton SO17 1BJ, United Kingdom}
\affiliation{IQFR-CSIC, Serrano 119, 28006 Madrid, Spain}

\begin{abstract}
Generating, controlling, and sensing strong magnetic fields at ever shorter time and length scales is important for both fundamental solid-state physics and technological applications such as magnetic data recording. Here, we propose a scheme for producing strong ultrashort magnetic pulses localized at the nanoscale. We show that a bimetallic nanoring illuminated by femtosecond laser pulses responds with transient thermoelectric currents of picosecond duration, which in turn induce Tesla-scale magnetic fields in the ring cavity. Our method provides a practical way of generating intense nanoscale magnetic fields with great potential for materials characterization, terahertz radiation generation, and  data storage applications.
\end{abstract}

\maketitle

\begin{figure}
\includegraphics[width=0.7\textwidth]{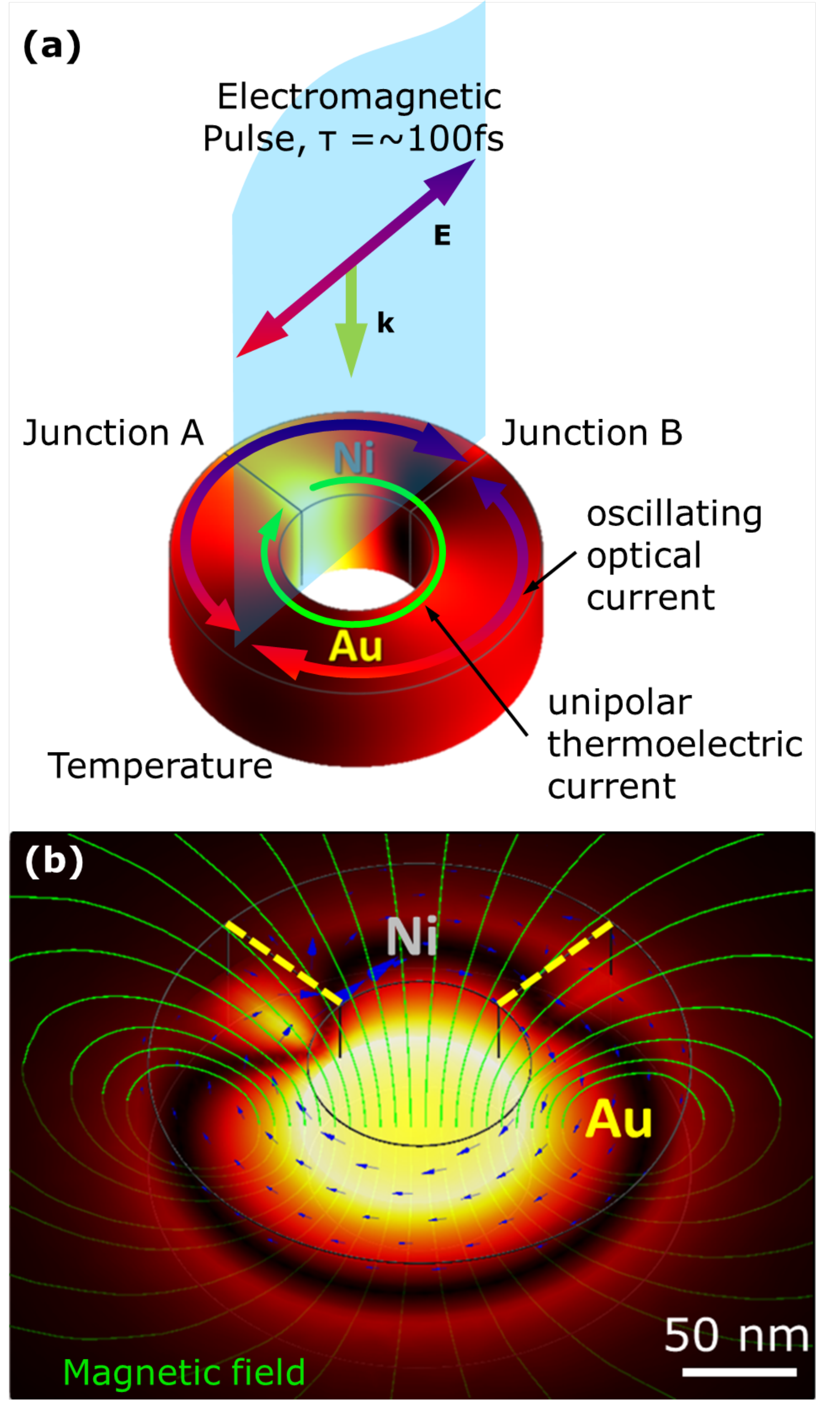}
\caption{{\bf Schematic description of our nanoscale ultrashort strong magnetic-field source.} {\bf (a)} A femtosecond near-infrared laser pulse heats a bimetallic ring consisting of one quadrant of nickel and three quadrants of gold. At the plasmonic resonance of the ring, a dipole current oscillating at the incident light frequency is excited (blue/red arrows), leading to absorption and subsequent conduction-electron heating. A strong, non-oscillating thermoelectric current is then produced in this short-circuited thermocouple, driven by the conduction-electron temperature difference ($\sim\,$thousands of degrees) between the cold and hot Au/Ni junctions. {\bf (b)} The thermoelectric ring current generates a strong magnetic field normal to the plane of the ring. The intensity of the magnetic-field component perpendicular to the ring is represented by the density plot, whereas magnetic-field lines are shown in green.}
\label{Fig1}
\end{figure}

\section{Introduction}

The study of ultrashort magnetic phenomena has been largely driven by their application to magnetic storage technologies, which are rapidly evolving towards the nanoscale and sub-picosecond regimes, giving rise to significant advances in the understanding of magnetization dynamics \cite{KKR10}. Currently, no easily accessible method is available to generate intense sub-picosecond magnetic pulses localized at the nanoscale. Short electron pulses can be tightly focused and are accompanied by magnetic fields of a few Teslas, but they require access to electron accelerators \cite{TSK04}. Transient magnetization has been also studied via the inverse Faraday effect in cases for which absorption of circularly polarized light leads to a prevailing spin species \cite{HMF1971,ZMH94,KKU05,HKK06}, although this technique is only applicable to highly absorbing materials exhibiting strong spin-orbit coupling, with spatial resolution limited by the light focal spot. Here, we propose a radically new approach towards ultrafast magnetism with nanoscale resolution relying on optically driven generation of magnetic fields.

In this work, we show that bimetallic nanorings can act as nanoscale sources of intense ultrashort magnetic pulses. We rely on the enhanced light absorption associated with the plasmons of metallic rings to generate transient thermoelectric currents that in turn produce sub-picosecond pulses with magnetic fields as high as a few tenths of a Tesla in the vicinity of the rings. The ability to generate strong magnetic fields localized on the nanoscale is of interest for elucidating spin and magnetization dynamics at sub-picosecond time and nanometer length scales \cite{KKR10}, and it holds great potential for materials characterization, terahertz radiation generation, and magnetic recording.

\section{Results and Discussion}

A scheme of our magnetic source is shown in Fig.\ \ref{Fig1}. A $100\,$fs laser pulse heats a bimetalic ring consisting of three quadrants of gold and one quadrant of nickel. This choice of materials is a compromise between their thermoelectric response and their electrical resistance in order to optimize the resulting magnetic field. The ring is $100\,$nm thick, has inner (outer) diameter of $70\,$nm ($170\,$nm), and is embedded in glass. The incident pulse is polarized along one of the Au/Ni junctions (B in Fig.\ \ref{Fig1}a) and is centered at a wavelength of $920\,$nm, which corresponds to the lowest-order dipole plasmon resonance of the ring. In a homogeneous gold ring, symmetry would lead to resonant absorption at two diametrically opposed regions. However, the asymmetry of the bimetallic ring polarization leads to stronger absorption near one of the Au/Ni junctions (A in Fig.\ \ref{Fig1}a). This nonuniform absorption produces a steep gradient of the electron temperature along the ring, and consequently, a net displacement of charge carriers occurs from hotter to colder regions, giving rise to a Seebeck electromotive force. In the Au/Ni thermocouple charge carriers in gold and nickel are electrons and holes, respectively, so that the resulting thermoelectric current contributes along the same direction on either side of each metal junction \cite{LZC08}. Femtosecond pulses give rise to temperature differences of several thousand degrees in the conduction electron gas. This results in a strong thermoelectric current before thermalization into the atomic lattice takes place within a few picoseconds. Finally, this transient circular current generates a magnetic field, mainly perpendicular to the plane of the ring (green lines in Fig.\ \ref{Fig1}b). We solve the underlying nonlinear heat-transfer/thermoelectric/electromagnetic problem within the two-temperature model \cite{LZC08,KKR10} using a commercial finite-elements tool (COMSOL), and we supplement these results with a semi-analytical model that correctly predicts the magnitude of the generated field and its duration (see Appendix\ \ref{semianalytical} for more details).

\begin{figure}
\includegraphics[width=0.7\textwidth]{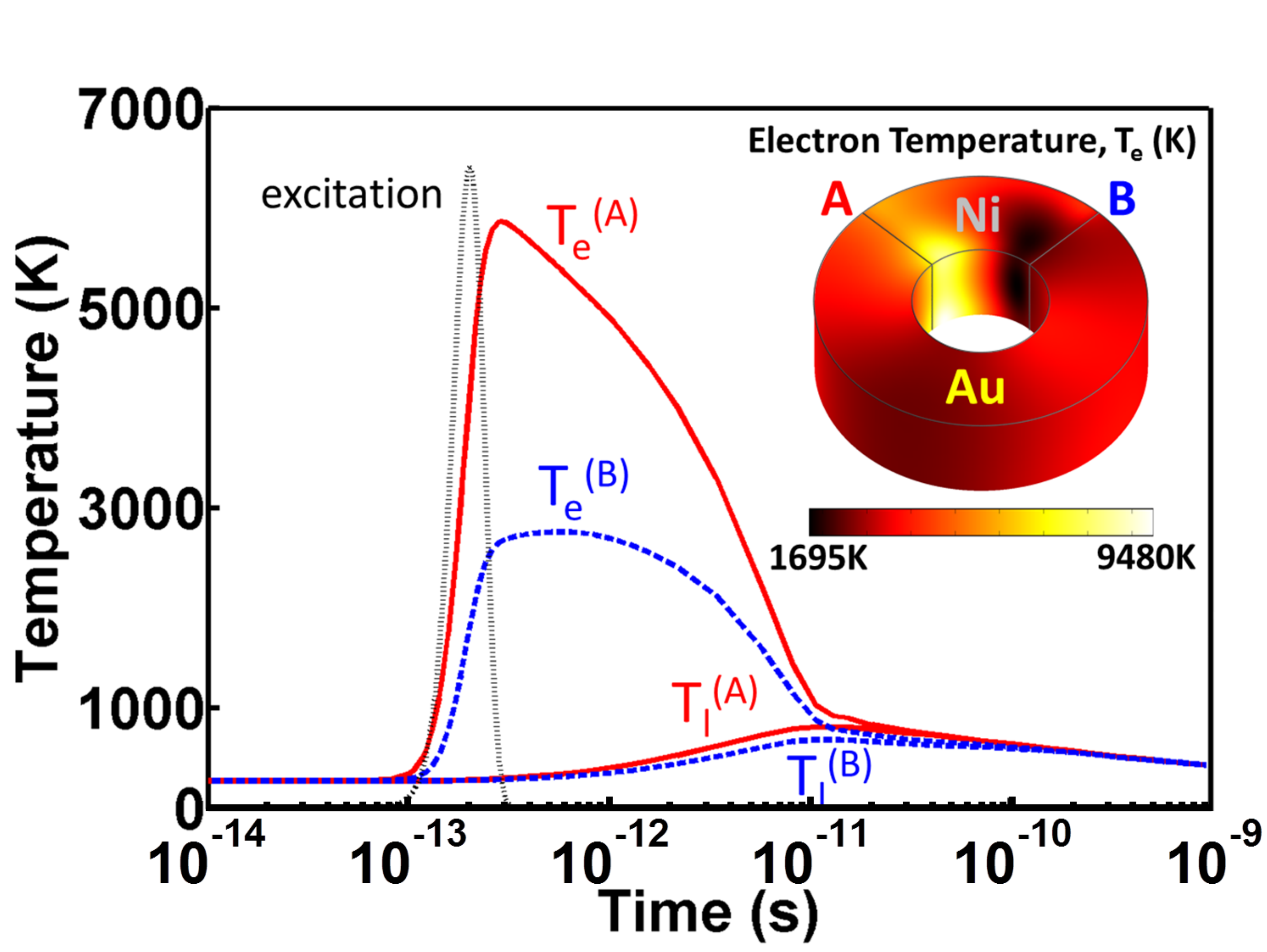}
\caption{{\bf Temporal temperature profiles upon irradiation by a light pulse.} Electron ($\Te$) and lattice ($\Tl$) temperatures at the hot (solid red, A) and cold (dashed blue, B) junctions of the ring. The pulse fluence is $11.3\,$J/m$^2$ and evolves as shown by the dotted curve. A snapshot of the electron temperature on the ring surface is shown in the inset at the time corresponding to the maximum of $\Te$ in the hot junction A.}
\label{Fig2}
\end{figure}

The time evolution of the conduction-electron ($\Te$) and lattice ($\Tl$) temperatures at the two Au/Ni junctions is shown in Fig.\ \ref{Fig2} for an incident pulse fluence of $11.3\,$J/m$^2$, which transfers $\sim1.4\,$pJ to the ring (absorption cross-section $\sim0.12\,\mu$m$^2$, see Appendix\ \ref{opticalheating}). Following an initial excitation with the ultrafast optical pulse, the electron-gas temperature rapidly rises at both junctions, reaching peak values shortly after the maximum of the incident pulse. The nickel section of the ring produces a considerable temperature difference between the two junctions: electrons at the hot junction (A) reach a peak temperature of $\sim6000\,$K, which is approximately twice the temperature of the cold junction electrons (B). The peak temperature at B occurs nearly 1\,ps after the maximum of A as a result of hot-electron diffusion. Over this short time interval, the absorbed energy remains almost entirely in the electron gas, while thermalization of the gold and nickel lattices, initially at $300\,$K, takes place over a time scale of several picoseconds. The lattice and electron temperatures become nearly identical $\sim10\,$ps after irradiation, when both junctions are at $\sim600\,$K, well below the melting temperature of the ring materials. Finally, the ring cools back to room temperature after a few nanoseconds due to diffusion through the surrounding glass.

The decay from maximum heating reveals three different time scales: a fast decay ($\sim\,1$ps) produced by thermal diffusion within the electron gases of the two metals; an intermediate decay ($\sim10\,$ps) that can be ascribed to thermalization of the electrons into the lattice via electron-phonon scattering; and a slow decay ($\sim$nanoseconds) arising from thermalization into the surrounding glass. This picture is consistent with the temperature dependence of the involved materials parameters (see Methods).

\begin{figure*}[htpb]
\includegraphics[width=0.95\textwidth]{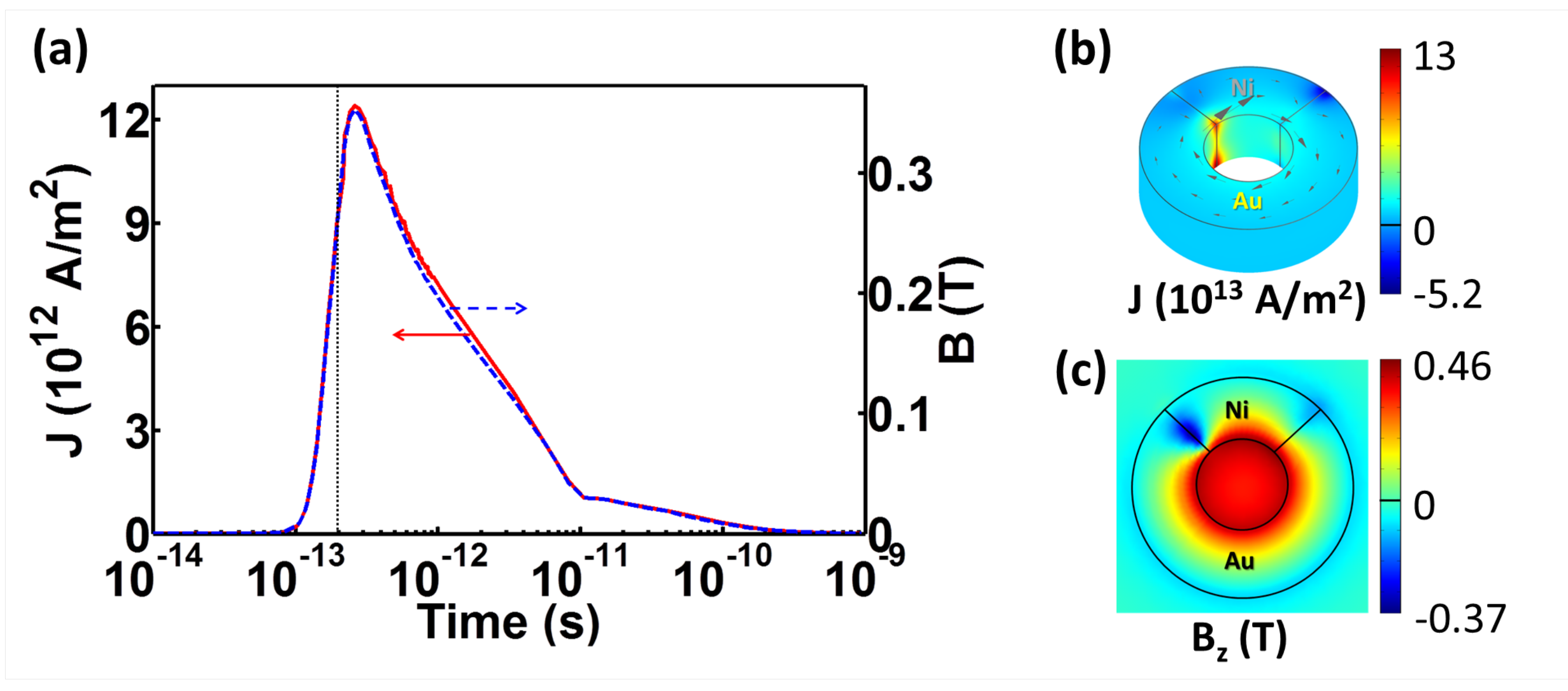}
\caption{{\bf Temporal thermoelectric current and magnetic field profiles.} {\bf (a)} We show the time evolution of the azimuthial component of the current through junction A (dashed curve, left scale) and the magnetic induction averaged over the ring hole (solid curve, right scale). The dotted vertical line indicates the maximum of the light pulse intensity. {\bf (b,c)} Snapshots of the instantaneous current on the ring surface (b) and the axial component of the magnetic induction (c), both at the time of the maximum of the curves in (a). The irradiation conditions are the same as in Fig.\ \ref{Fig2}.}
\label{Fig3}
\end{figure*}

The temperature gradient between two junctions of metals with different Seebeck coefficients produces an electromotive force that drives a thermoelectric current along the ring circumference. For uniformly heated junctions, this effect is quantified through a potential between them $S\Delta T$, proportional to the temperature difference $\Delta T$. The Seebeck coefficient $S$ depends on the nature of the material between the two junctions, as well as on temperature (see Appendix\ \ref{ethparam}). Because of the high electron temperatures involved, the thermoelectric current density in the confined ring geometry reaches values as high as $\sim10^{13}\,$A/m$^2$. Figure\ \ref{Fig3}a (left scale) illustrates the temporal evolution of the current density through junction A, which follows quite closely the temperature difference, with a maximum occurring $\sim100\,$fs after the peak of the excitation pulse. The current distribution on the ring surface (Fig.\ \ref{Fig3}b) reveals a dominant azimuthal component, except near the hot junction, where parasitic short-range loops are produced.

\begin{figure*}
\includegraphics[width=0.8\textwidth]{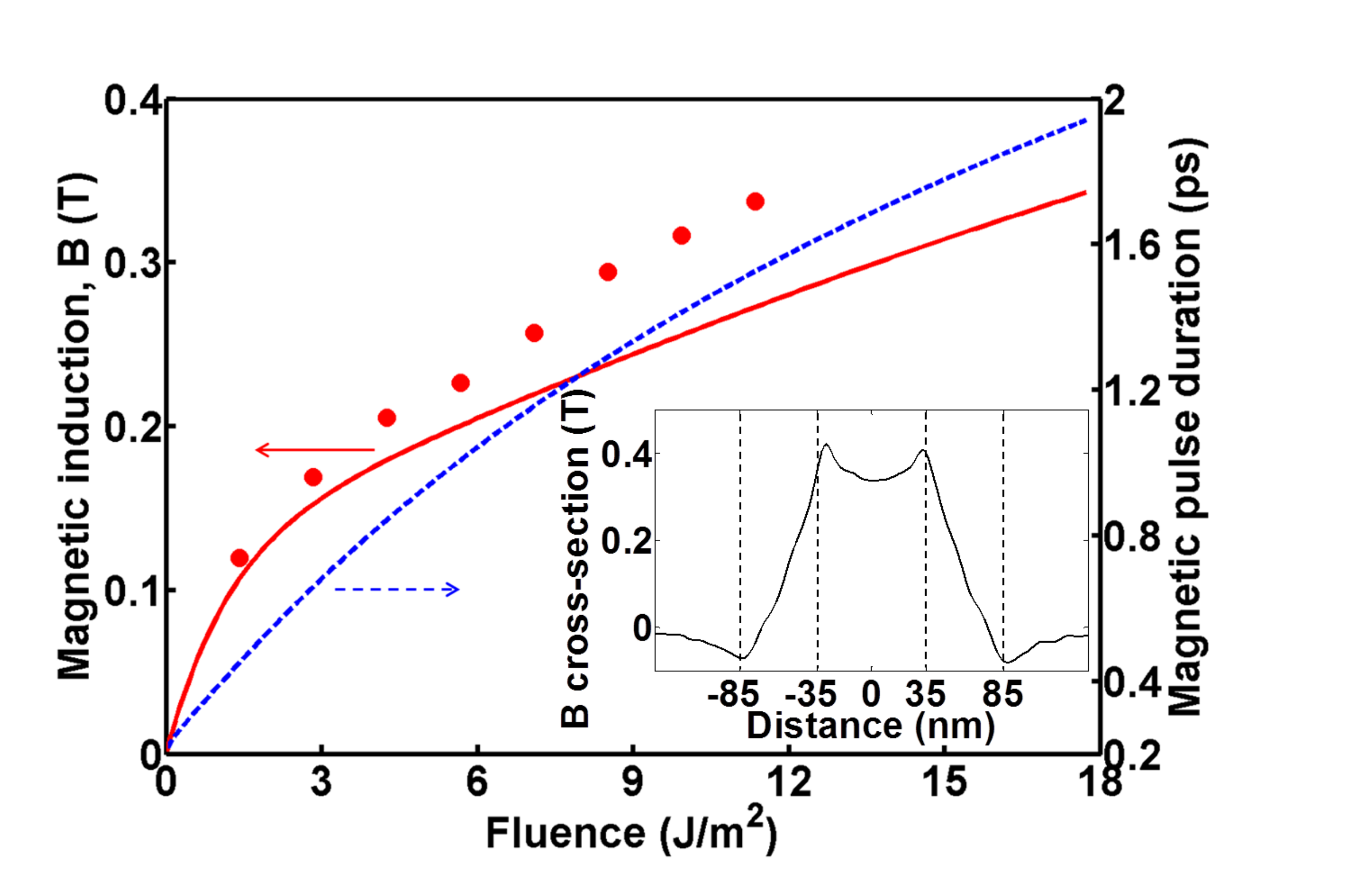}
\caption{{\bf Magnetic-field strength and magnetic-pulse duration as a function of light-pulse fluence.} We represent the calculated maximum of the magnetic induction (left vertical scale) as a function of incident pulse fluence from full numerical simulations (symbols) and from the semi-analytical model described in Appendix\ \ref{semianalytical} (solid curve, Eq.\ (\ref{B})). The FWHM duration of the magnetic pulse (right vertical scale) is estimated from the analytical model (dashed curve, Eq.\ (\ref{duration})). The inset shows a characteristic profile of the magnetic induction across the ring.}
\label{Fig4}
\end{figure*}

These strong currents give rise to a large magnetic induction ($>0.35\,$T peak value) of similar temporal profile (Fig.\ \ref{Fig3}, right scale). The induction is localized inside the ring cavity (Fig.\ \ref{Fig3}c), where it is rather uniform (Fig.\ \ref{Fig4}, inset), with the exception of a depletion caused by the noted loop currents in the hot Au/Ni junction. The magnetic pulse has a duraction $\Delta t_{\rm mag}\sim1.8\,$ps FWHM, mainly controlled by electron diffusion and thermalization into the lattice (see Methods).

The peak magnetic induction grows monotonically with light-pulse fluence (Fig.\ \ref{Fig4}), exhibiting a slightly sublinear behavior as a result of the increase in electron heat capacity with temperature \cite{LZC08}, although the net effect involves a complex interplay between the temperature dependence of the materials thermal and electrical parameters (see Methods).

We have formulated a simple model that yields semi-analytical expressions for the peak magnetic induction (Eq.\ (\ref{B})) and the magnetic pulse duration (Eq.\ (\ref{duration})), as discussed in Appendix\ \ref{semianalytical}. The model is based upon the neglect of thermal diffusion in the lattice and it assumes that the optical pulse acts as a sudden impulse. This produces results in excellent agreement with full numerical simulations, as shown in Fig.\ \ref{Fig4}, thus providing a simple tool that can assist in the design of rings optimized in size and composition to yield the desired levels of magnetic pulse intensity and duration. In particular, the model predicts a magnetic field duration that increases with optical pulse fluence, which essentially reflects the decrease in electron-phonon coupling and the increase in the heat capacity associated with the ever larger electron temperature that is reached with higher pumping intensities (see Appendix\ \ref{ethparam1}).

\begin{figure}
\begin{center}
\includegraphics[width=170mm,angle=0,clip]{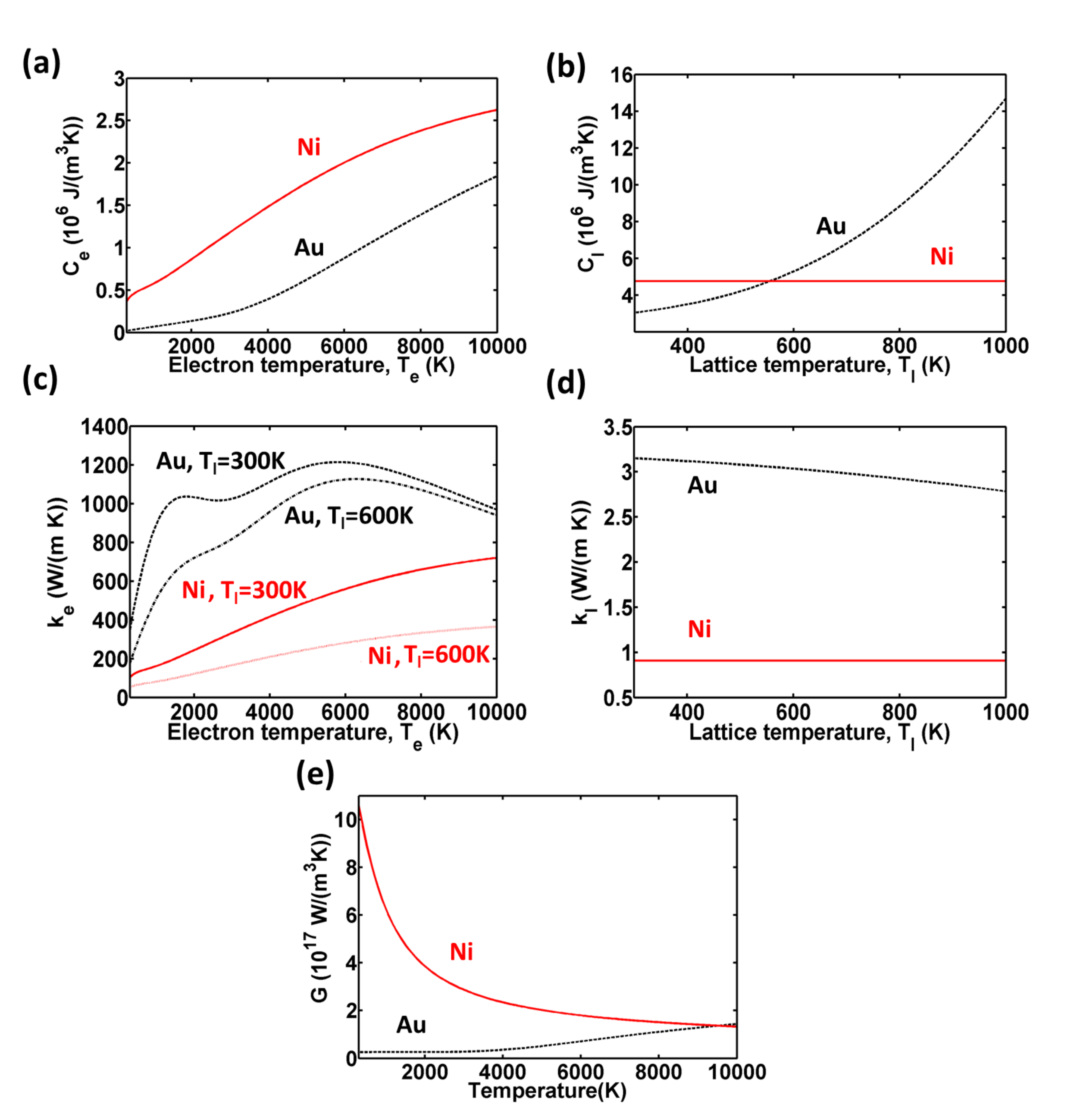}
\caption{{\bf Thermal parameters used in the description of heat diffusion for Ni and Au.} {\bf (a,b)} Electronic \cite{LZC08} and lattice \cite{HZC09,L05} heat capacities, $\Ce$ and $\Cl$, respectively. {\bf (c,d)} Electronic \cite{AM1976,CLB05,IZ03} and lattice \cite{HZC09,L05} thermal conductivities, $\ke$ and $\kl$, respectively. {\bf (e)}  Electron-lattice coupling coefficient \cite{LZC08}, $G$.}
\label{FigS1}
\end{center}
\end{figure}

\begin{figure}
\begin{center}
\includegraphics[width=170mm,angle=0,clip]{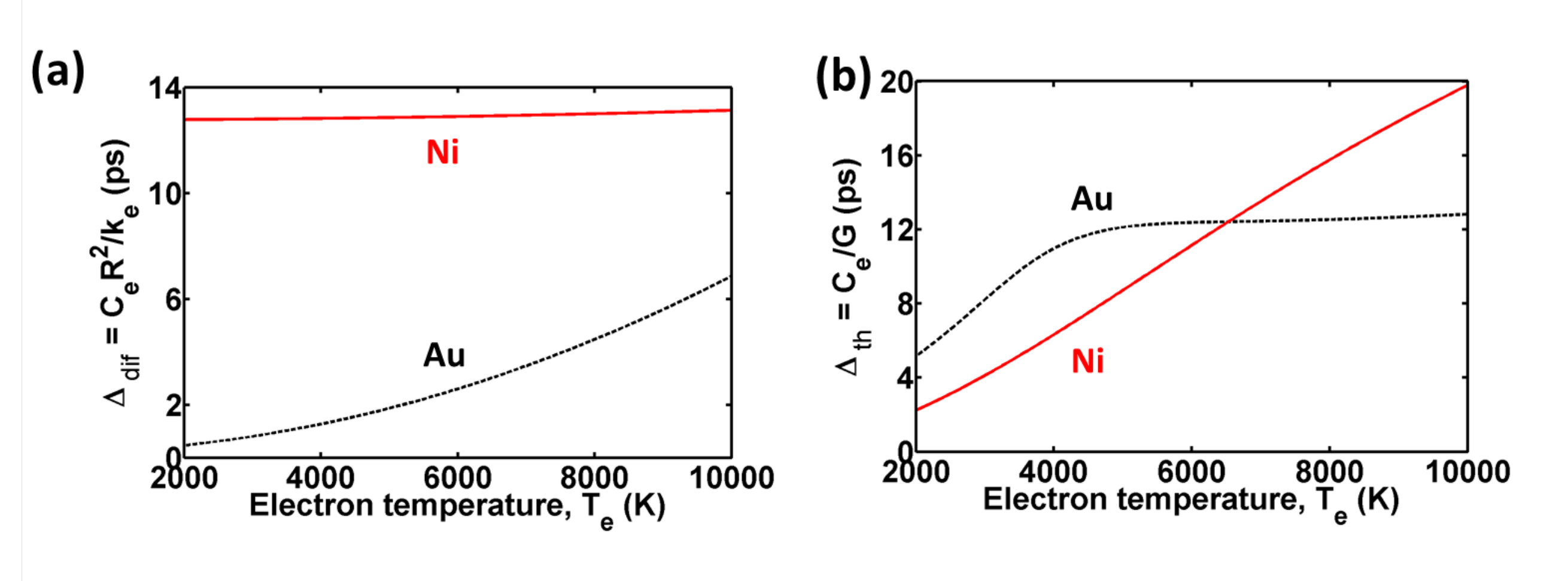}
\caption{{\bf Diffusion and thermalization.} Estimates of the local (a) diffusion and (b) thermalization times based upon the coefficients of Fig.\ \ref{FigS1} and their involvement in Eqs.\ (\ref{heat}).}
\label{FigS2}
\end{center}
\end{figure}

\begin{figure}
\begin{center}
\includegraphics[width=170mm,angle=0,clip]{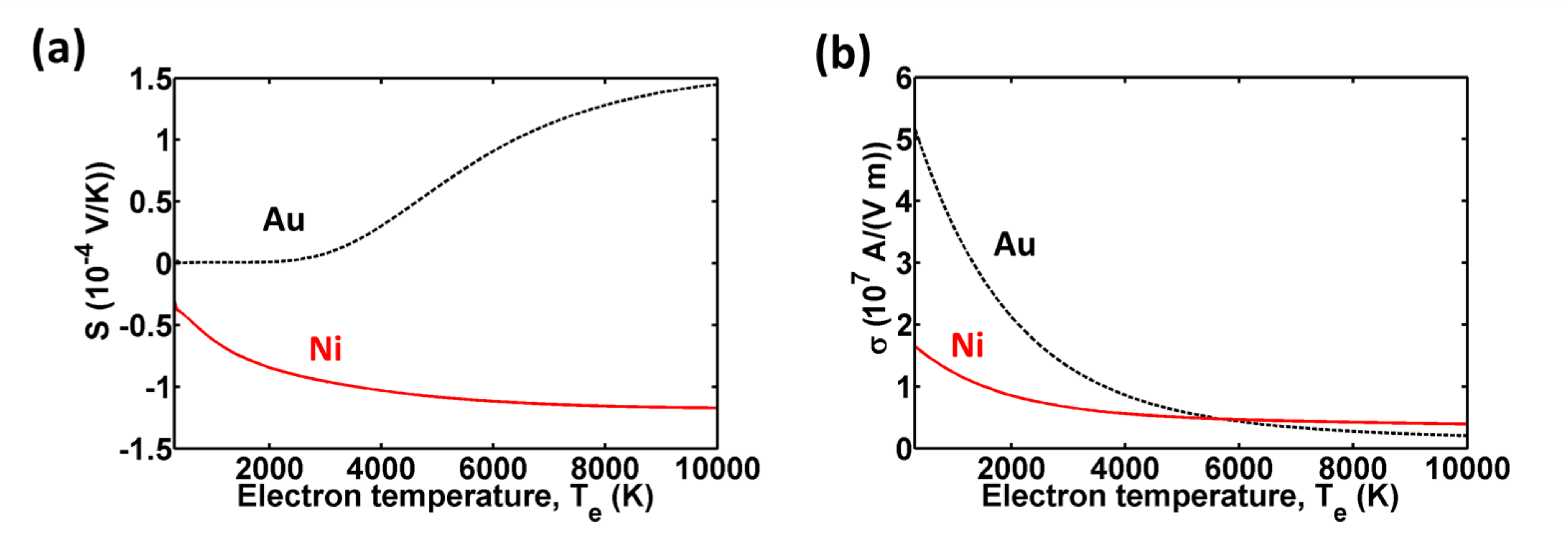}
\caption{{\bf Electrical and thermoelectrical properties.} Temperature dependence of (a) the Seebeck coefficient \cite{HBD10} and (b) the electrical conductivity \cite{HBD10,LZC08}.}
\label{FigS3}
\end{center}
\end{figure}

\begin{figure}
\begin{center}
\includegraphics[width=80mm,angle=0,clip]{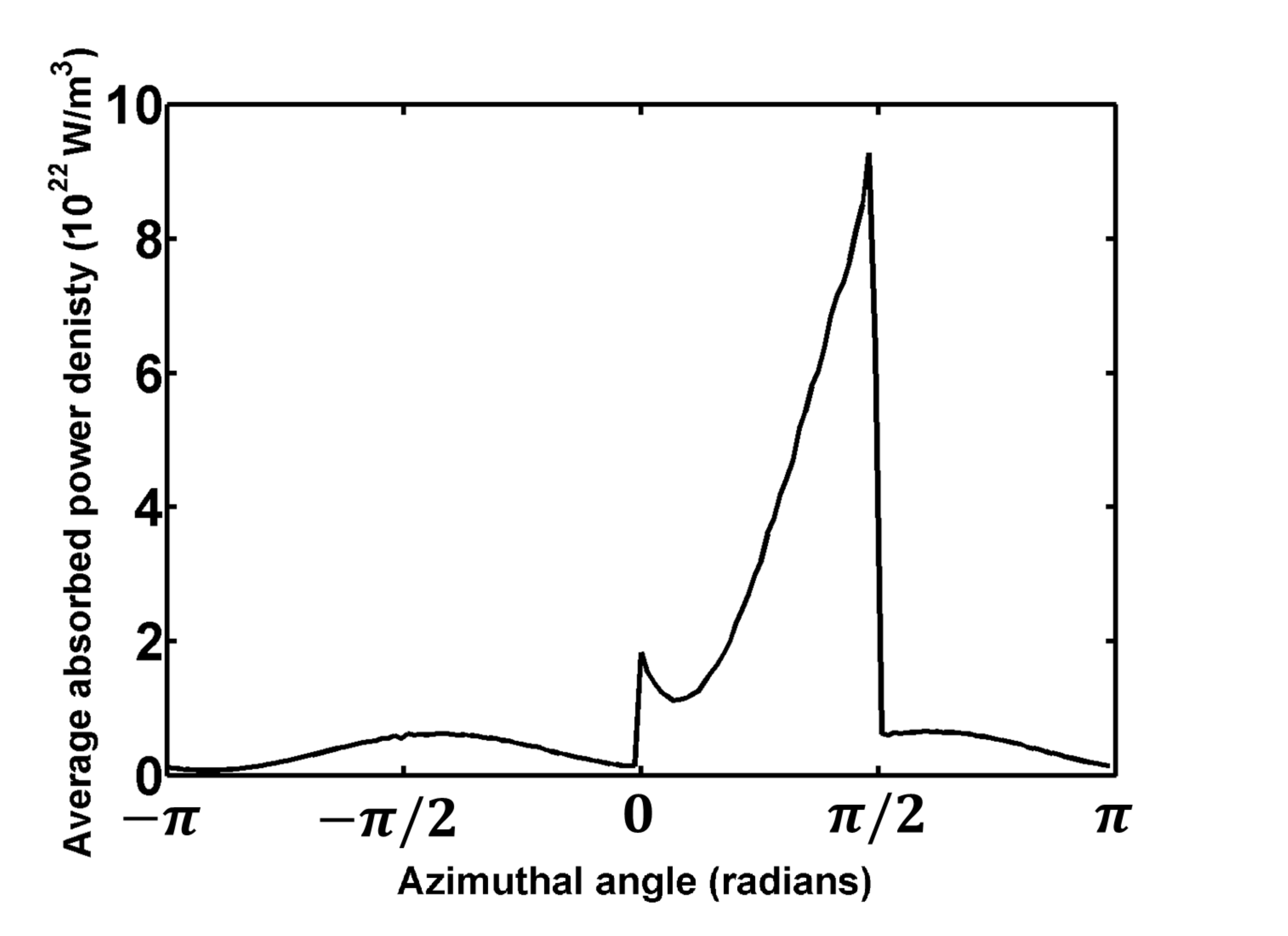}
\caption{{\bf Absorbed power density.} We plot the average of this quantity over the ring cross-section as a function of the azimuthal angle along the ring considered in Fig.\ \ref{Fig1} for illumination with a plane wave of intensity $10^{14}\,$W/m$^2$ at the resonance wavelength of 920\,nm. The external field is linearly polarized along the $0$ azimuthal-angle direction, which coincides with one of the Ni/Au junctions. The other junction is at angle $\pi/2$, as shown in Fig.\ \ref{Fig1}.}
\label{FigS4}
\end{center}
\end{figure}

\begin{figure}
\begin{center}
\includegraphics[width=80mm,angle=0,clip]{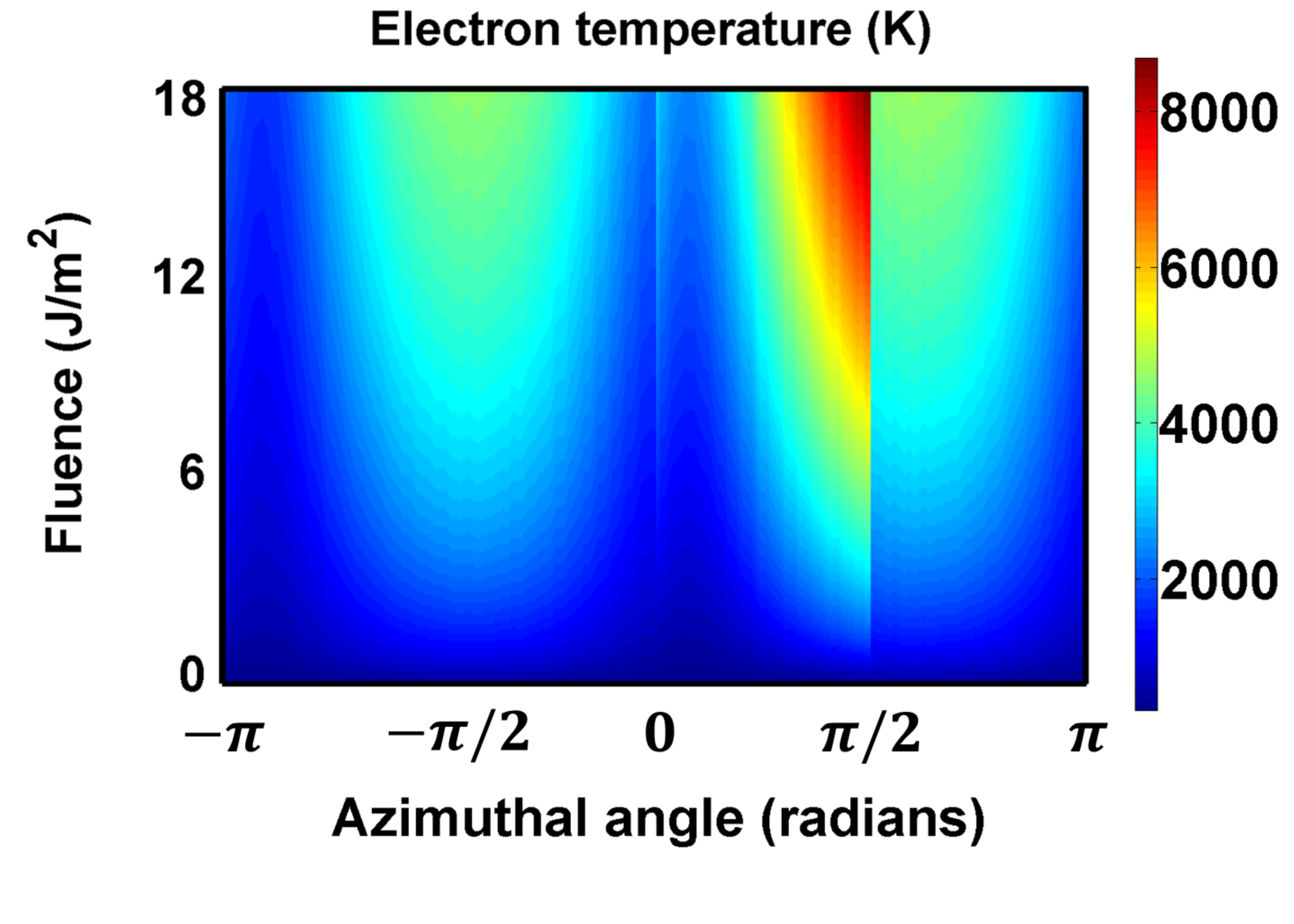}
\caption{Distribution of the electron temperature along the ring circumference right after light-pulse irradiation as a function of pulse fluence, according to the semi-analytical model described in Appendix\ \ref{semianalytical}.}
\label{FigS5}
\end{center}
\end{figure}

\begin{figure}
\begin{center}
\includegraphics[width=80mm,angle=0,clip]{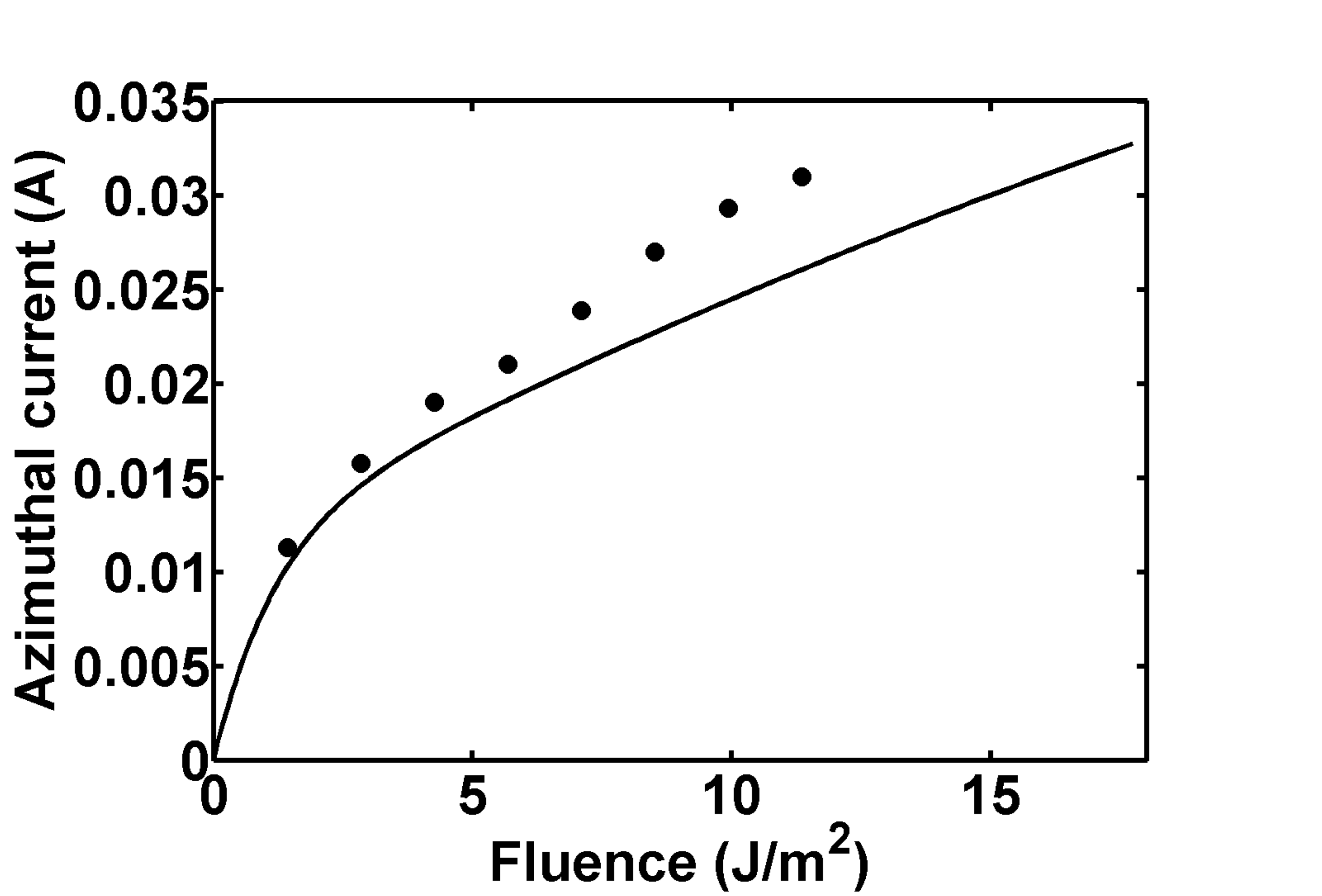}
\caption{Electrical current circulating around the ring right after light-pulse irradiation as a function of pulse fluence, obtained from the semi-analytical model (curve) and from full numerical simulations (symbols).}
\label{FigS6}
\end{center}
\end{figure}

\section{Conclusion and Perspectives}

Our results indicate that we can take advantage of the strong thermoelectric currents at the nanoscale in order to achieve very intense, transient, THz electromagnetic fields driven by pumping with vis-NIR light. This can be the basis of a new source of optically pumped THz radiation, whereby illuminated rings act as localized magnetic dipoles. In a different, important range of applications, illuminated bimetallic rings provide an interesting opportunity for nanoscale magnetic recording triggered by the intense magnetic field (strongly confined in the $<100\,$nm ring cavity) in combination with suitable magnetic grain sizes that can react to the short duration of the magnetic pulse \cite{TSK04}. As a practical consideration, the main limitation on the values of the achievable magnetic field comes from the breakdown current in the rings ($\sim10^{12}-10^{13}\,$A$/$m$^2$ for gold \cite{KMA09,YDM11}), as well as from possible melting of the metals \cite{ISY05}. However, the short duration of the optical irradiation, the moderate lattice temperatures involved, and the embedding of the ring in a rigid glass matrix should push the thresholds for melting and electrical breakdown well beyond the currents and temperatures considered here.

In conclusion, we have shown that a bimetallic nanoring illuminated by ultrafast laser pulses supports transient thermoelectric currents leading to magnetic pulses of sub-picosecond duration reaching a fraction of a Tesla and localized in the $<100\,$nm ring cavity. Our results can facilitate the study of ultrafast, nanoscale magnetic phenomena, and they hold great potential for applications in materials characterization, the generation of terahertz radiation, and magnetic recording.

\section*{ACKNOWLEDGMENT}

This work has been supported in part by the Engineering and Physical Sciences Research Council (U.K.), the Leverhulme Trust, the Royal Society, and the Spanish MEC (MAT2010-14885 and Consolider NanoLight.es).

\appendix

\section{Theoretical formalism}
\label{theory}

Our simulations of optically induced generation of magnetic fields mediated by thermoelectric currents in bimetallic rings involve solving a set of coupled equations for (1) the electromagnetic field intensity of the incident light, (2) the heat diffusion problem, (3) the resulting thermoelectric current, and (4) the magnetic field itself. In this work, we provide two levels of theory based upon (i) full numerical simulations and (ii) semi-analytical modeling. We use a commercial finite-element method (COMSOL) for the former, whereas a derivation of the latter is summarized below. But first, we discuss the relevant equations and approximations required to deal with each of the four noted subproblems.

{\bf (1) Light-pulse absorption by the ring.} Light absorption is calculated by solving Maxwell's equations for a Gaussian pulse of $\sim2\Delta t_{\rm pulse}=100\,$fs duration centered at a wavelength of $920\,$nm and incident on the ring along its axis with linear polarization as shown in Fig.\ \ref{Fig1}. Because the pulse duration is much longer than the light period (3.1\,fs), the absorbed power density (i.e., the position- and time-dependent power absorbed per unit volume inside the ring) is approximated as $p(\rb,t)=p(\rb)f(t)$, assuming that it follows the pulse temporal profile $f(t)=\exp[-t^2/(\Delta t_{\rm pulse})^2]$, multiplied by the peak absorption density $p(\rb)$, which we calculate for a plane wave with the same intensity as the pulse maximum. The dielectric functions of gold ($-33.56-1.93i$), nickel ($-15.47-26.03i$), and glass (2.1) are taken from tabulated optical data \cite{P1985}.

{\bf (2) Thermal diffusion and temperature distribution.} The absorbed power density $p(\rb,t)$ acts as the source of heat for the thermal diffusion problem. We adopt a two-temperature model \cite{LZC08}, in which the conduction-electron and atomic-lattice subsystems have their own temperatures $\Te(\rb,t)$ and $\Tl(\rb,t)$, respectively, each of them evolving over very different time scales. We assume that the heating produced through $p(\rb,t)$ is entirely transferred from the light to the conduction electrons, which reach local thermal equilibrium in a short time compared with the characteristic times for thermal diffusion and lattice thermalization (see below). The thermal diffusion equation is then simultaneously solved for the electronic and lattice subsystems, which are coupled through a term accounting for heat transfer from the former to the latter. We assume this term to be proportional to the local temperature difference $\Te-\Tl$, via an electron-phonon coupling coefficient $G$. This leads to the two coupled diffusion equations \cite{LZC08}
\begin{subequations}
\label{heat}
\begin{equation}
\label{heata}
\Ce\dot{\Te}=\nabla (\ke\nabla \Te) - G\,(\Te-\Tl)+p,
\end{equation}
\begin{equation}
\Cl\dot{\Tl}=\nabla (\kl\nabla \Tl) + G\,(\Te-\Tl),
\end{equation}
\end{subequations}
where $\Ce$ and $\Cl$ are the electron and lattice heat capacities, and $\ke$ and $\kl$ are the respective thermal conductivities. We use values of the temperature-dependent coefficients $C$, $\kappa$, and $G$ for Ni and Au compiled from Refs. \cite{LZC08,HZC09,L05,AM1976,CLB05,IZ03}, as discussed in Appendix\ \ref{ethparam1}. For simplicity, we ignore the effect of heterojunction thermal barriers, which should contribute to create higher temperature gradients and stronger magnetic fields.

{\bf (3) Thermoelectric current.} From the solution of Eqs.\ (\ref{heat}), we derive a thermoelectric electromotive force $S\nabla\Te$, where $S$ is the Seebeck coefficient. This allows us to write the thermoelectric current as
\begin{align}
\label{j}
\jb=\sigma\left(S\nabla\Te-\nabla V\right),
\end{align}
where $\sigma$ is the temperature-dependent electric conductivity and $V$ represents the potential displayed in response to the thermoelectric source. We use values for the temperature-dependent coefficients $S$ and $\sigma$ for Ni and Au calculated from Refs.\ \cite{HBD10,LZC08}, as explained in Appendix\ \ref{ethparam}. Because $\Te$ varies smoothly over hundreds of femtoseconds, involving characteristic frequencies below the mid-infrared, we can safely assume that the metals behave as good conductors, so that the thermoelectric current is quasi-stationary and the continuity equation reduces to the vanishing of $\nabla\cdot\jb$ (this is equivalent to neglecting self-inductance of the thermoelectric current). Inserting Eq.\ (\ref{j}) in this expression, we find
\begin{align}
\label{V}
\nabla\cdot\sigma\nabla V=\nabla\cdot\sigma S\nabla\Te,
\end{align}
which is formally equivalent to Poisson's equation with a local response function $\sigma$ and a source term $-\nabla\cdot\sigma S\nabla\Te$. Incidentally, by setting $\sigma=0$ outside the ring, Eq.\ (\ref{V}) guarantees the vanishing of the normal current at the ring boundaries. We obtain $V$ by solving Eq.\ (\ref{V}), and this is in turn inserted into Eq.\ (\ref{j}) to yield $\jb$.

{\bf (4) Magnetic induction.} Finally, we use the electric current $\jb$ (Eq.\ (\ref{j})) to find the resulting magnetic induction from \cite{J99}
\begin{equation}
{\bf B}(\rb,t)=\frac{\mu_0}{4\pi}\int d^3\rb'\,\frac{\jb\times(\rb-\rb')}{|\rb-\rb'|^3},
\nonumber
\end{equation}
where the integral is extended over the volume of the ring.

\section{Semi-analytical model}
\label{semianalytical}

A simple model that predicts the order of magnitude of the generated magnetic field, the current circulating around the ring, and the duration of the magnetic pulse can be formulated under the following approximations:

{\bf (i) Sudden pulse excitation.} Upon inspection of Eq.\ (\ref{heata}), we expect to have characteristic diffusion and thermalization times given by $\Delta t_{\rm dif}=R^2\Ce/\ke$ and $\Delta t_{\rm th}=\Ce/G$, respectively, where $R$ is the average ring radius. For the high temperatures ($\Te>2000\,$K) at which  a large magnetic field is generated, we have $\Delta t_{\rm dif}>0.5\,$ps and $\Delta t_{\rm th}>2\,$ps (see Appendix\ \ref{opticalheating}), in front of which we neglect $\Delta t_{\rm pulse}=50\,$fs, and thus we consider the pulse to deliver an instantaneous impulse quantified by an absorbed energy density $q=\sqrt{\pi}\Delta t_{\rm pulse}p(\rb)$. This determines the maximum local electron temperature $\Te$ right after pulse irradiation by imposing $q=\int_{T_0}^{\Te}\Ce(T)\, dT$, where $T_0$ is the initial room temperature before irradiation.

{\bf (ii) Neglect of the lattice.} Upon examination of the physical parameters entering Eqs.\ (\ref{heat}) (see Appendices\ \ref{ethparam1} and \ref{ethparam}), we observe that $\kl\ll\ke$ and $\Cl\gg\Ce$, that is, the lattice takes much more heat to increase its temperate than the electron gas and it conducts heat more poorly. This explains why $\Tl\ll\Te$ during the bulk of the magnetic pulse duration, so we can safely ignore the lattice and reduce Eqs.\ (\ref{heat}) to $\Ce\dot{\Te}=\nabla (\ke\nabla \Te) - G\,\Te$, with the initial condition for $\Te(t=0)$ derived above.

{\bf (iii) Average over ring cross-section.} We can further assume that the temperature distribution can be well represented if we replace it by its azimuthal-angle-dependent average over the radial and vertical directions (i.e., the average over the ring cross-section). We then focus on the azimuthal component of the current $j$, which we can take as uniform along the ring circumference as a consequence of the vanishing of $\nabla\cdot\jb$. Noticing that the potential $V$ must reach the same value when $\nabla V$ is integrated along a whole ring circumference, we find from Eq.\ (\ref{j}) that $j\int d\varphi/\sigma=(1/R)\int S (\partial\Te/\partial\varphi)d\varphi$, where we have approximated the azimuthal component of $\nabla$ as $(1/R)\partial/\partial\varphi$. Finally, we can carry out the integral in the right-hand side of this equation from junction A to B through Au and from B to A through Ni (see Fig.\ \ref{Fig2}a). This produces the expression
\begin{align}
B=\frac{\mu_0}{4\pi}\,\beta\,\frac{\int_{T_B}^{T_A}dT\,[S_{\rm Au}(T)-S_{\rm Ni}(T)]}{\int d\varphi/\sigma}
\label{B}
\end{align}
for the $z$ component of the magnetic induction as at the ring center, where $\beta=(1/R)\int d\rb\,\hat{\bf z\cdot}(\hat{\varphi}\times\rb)/r^3$ is a dimensionless geometrical factor ($\beta\approx V/R^3$, where $V$ is the ring volume).

Following the discussion in point (i), we estimate the pulse duration from the diffusion and thermalization times as
\begin{equation}
\Delta t_{\rm mag}=\left\langle\frac{1}{1/\Delta t_{\rm dif}+1/\Delta t_{\rm th}}\right\rangle,
\label{duration}
\end{equation}
where the brackets stand for the average along the ring circumference, which needs to be taken due to the strong dependence of $\Delta t_{\rm dif}$ and $\Delta t_{\rm th}$ on the local electron temperature.
We show in Fig.\ \ref{Fig4} the pulse duration evaluated from Eq.\ (\ref{duration}) right after irradiation.

\section{Thermal parameters}
\section{ethparam1}

The parameters describing heat diffusion in both the electronic and lattice subsystems of Ni and Au are plotted in Fig.\ \ref{FigS1}, along with the coefficient for electron-lattice thermalization. These parameters depend on the electron temperature. Additionally, the electron thermal conductance depends on the lattice temperature.

The temperature-dependent electron heat capacities are taken from Ref.\ \cite{LZC08} (Fig.\ \ref{FigS1}a), whereas the lattice heat capacity for Au is taken from Ref.\ \cite{HZC09} (Fig.\ \ref{FigS1}b). However, there is little information available on the temperature dependence of the lattice heat capacity for Ni, and since the lattice temperature is only mildly affecting our final results and the range of lattice temperatures is limited to $\sim300-600\,$K, we have approximated it by a constant value measured at room temperature \cite{L05}.

We approximate the electron thermal conductivity  (Fig.\ \ref{FigS1}c) from the Drude model as \cite{AM1976}
\begin{equation}
\ke=\frac{1}{3}\Ce v_F^2\tau,
\nonumber
\end{equation}
where $v_F$ is the Fermi velocity ($1.40\times10^6\,$m/s in Au \cite{AM1976} and $0.28\times10^6\,$m/s in Ni \cite{PAH98}) and $\tau$ is the electron scattering rate. The latter is approximated as \cite{CLB05}
\begin{equation}
\tau=\frac{1}{A_e\Te^2+B_l\Tl},
\label{tau}
\end{equation}
where $A_e=1.2\times10^7$ and $B_l=1.23\times10^{11}$ for Au \cite{CLB05} and $A_e=1.4\times10^6$ and $B_l=1.624\times10^{13}$ for Ni \cite{IZ03}, all in SI units with the temperatures in Kelvin. The lattice thermal conductivity (Fig.\ \ref{FigS1}d) is taken from Ref.\ \cite{HZC09} for Au and approximated as a constant for Ni \cite{L05} because of the reasons discussed above. Finally, the electron-lattice coupling coefficient $G$ (Fig.\ \ref{FigS1}e) is taken from Ref.\ \cite{LZC08}.

From the form of Eqs.\ (\ref{heat}), we can extract two characteristic times $\Delta t_{\rm dif}=R^2\Ce/\ke$ and $\Delta t_{\rm th}=\Ce/G$ associated with heat diffusion and thermalization, respectively. Here, $R=60\,$nm is the average radius of the ring. We plot these times in Fig.\ \ref{FigS2} as a function of electron temperature using the parameters of Fig.\ \ref{FigS1}. Both of these types of processes are thus expected to have similar influence on the temporal decay of the high temperature gradient following light pulse irradiation.

\section{Electrical and thermoelectrical parameters}
\label{ethparam}

The Seebeck coefficient $S$ and the electrical conductivity $\sigma$ are calculated from the electronic density of states (taken from \cite{LZC08} for Au and Ni), conveniently weighted by the Fermi-Dirac distribution for each temperature $\Te$, following the prescription described in Ref.\ \cite{HBD10}. Both $S$ and $\sigma$ exhibit a pronounced electron-temperature dependence, as shown in Fig.\ \ref{FigS3}.

\section{Optical heating}
\label{opticalheating}

We calculate the absorption power density $p(\rb)$ in the ring upon illumination by a plane wave of intensity $10^{14}\,$W/m$^2$, linearly polarized with the electric field along the $x$ axis (i.e., azimuthal angle 0, coinciding with one of the Au/Ni junctions). A commercial finite-difference method (COMSOL) is used for this calculation. The absorption density is simply scaled with the temporal profile of the light pulse, as explained in Appendix \ref{theory}. The entire $\rb$ dependence of $p(\rb)$ is used in our full numerical simulations, whereas we average this quantity over the cross section of the ring as an input for the semi-analytical model. The averaged absorption density exhibits pronounced maxima at the Au/Ni junctions (Fig.\ \ref{FigS4}) and clearly reveals the hot junction at an azimuthal angle of $\pi/2$ (i.e., where the current at the optical frequency is maximum upon excitation of a dipole plasmon along the $x$ axis). The overall absorption is quantified by an effective cross-section of $\sim1.2\,\mu$m$^2$.

The distribution of the electron temperature along the ring circumference is shown in Fig.\ \ref{FigS5} as a function of pulse fluence, as obtained from the semi-analytical model, averaged over the ring-cross section. A nonlinear dependence on fluence is clearly observed.

\section{Thermoelectrical current}

We show in Fig.\ \ref{FigS6} the thermoelectric current obtained from the semi-analytical model (solid curve), as compared with full numerical simulations (symbols). In the semi-analytical model, the generated magnetic field (Fig.\ \ref{Fig4}) follows exactly the same profile as the current, whereas the full simulation shows minor differences between the two profiles due to the finite spatial extension of the current over the cross section of the ring.

The current density (e.g., $\sim5\times10^{12}\,$A/m$^2$ for a fluence of 2\,J/m$^2$) is above the threshold for materials damage ($\sim10^{12}\,$A/m$^2$ from measurements in nanowires under DC conditions \cite{KMA09}). However our ring is only exposed to it during a short interval, well below the characteristic time scale for structural changes in the metals. Embedding of the ring in a hard,transparent material can be a good strategy for increasing the damage threshold.

%\bibliographystyle{apsrev}
%\bibliography{../../bibtex/refs}

\end{document}